\documentclass[twoside,12pt]{article}
\usepackage{epsfig}

\def\Journal#1#2#3#4{{#1} {#2} (#4) #3 }
\def\NCA{{\em Nuovo Cimento} A}

\def\NPA{{\em Nucl. Phys.} A}

\def\PRO{{\em Prog. Theor. Phys.}}
\def\NPB{{\em Nucl. Phys.} B}

\def\PLB{{\em Phys. Lett.} B}

\def\PL{{\em Phys. Lett.}}
\def\PRL{\em Phys. Rev. Lett.}
\def\PREV{\em Phys. Rev.}
\def\PREP{\em Phys. Rep.}

\def\PRD{{\em Phys. Rev.} D}
\def\PRC{{\em Phys. Rev.} C}

\def\ANNP{\em Ann. Phys. (N.Y.)}
\def\RMP{{\em Rev. Mod. Phys.}}

\newcommand{\be}{\begin{equation}}
\newcommand{\ee}{\end{equation}}
\newcommand{\bea}{\begin{eqnarray}}
\newcommand{\eea}{\end{eqnarray}}
\newcommand{\nn}{\nonumber}
\begin{document}

\title{
Introduction to Sterile Neutrinos
}

\author{Raymond R. Volkas
\\
School of Physics,\\ Research Centre for High Energy Physics,\\
The University of Melbourne, Victoria, 3010\\
Australia
}

\maketitle

\begin{abstract}
Model-building issues raised by the prospect of light sterile
neutrinos are discussed in a pedagogical way. 
I first review
the na\"{\i}ve proposal that sterile neutrinos be
identified with ``right handed neutrinos''. A critical
discussion of the
simple expedient of adding three gauge singlet fermions to
the usual minimal standard model matter content is 
followed by an examination of right handed neutrinos
in extended theories. I introduce the terminology of
``fully sterile'' and ``weakly sterile'' to classify
varieties usually conflated under the sterile 
neutrino banner.
After introducing the concepts
of ``technical naturalness'' and plain ``naturalness'', the
unbearable lightness of being a sterile neutrino is confronted.
This problem is used to motivate mirror neutrinos, whose
connection with pairwise maximal
mixing is emphasised. Some brief remarks about phenomenology
are made throughout. The impossibility of identifying the sole
sterile neutrino of the currently favoured $2 + 2$ and $3 + 1$
phenomenological constructs 
as a lone gauge singlet fermion added to the minimal
standard model is explained.
Finally, I remark on the beauty and subtlety of light sterile 
neutrino cosmology. 
\end{abstract}
\section{Introduction}

The discovery of sterile neutrinos would rank in importance no lower than the 
discoveries of charm, bottom and tau. The role of charm in the theory
of elementary particle interactions was presaged by Glashow, Iliopoulos
and Maiani (GIM): it was needed to remove tree-level flavour changing
neutral currents \cite{GIM}. Similarly, Kobayashi and Maskawa \cite{KM}
had anticipated
the need for a third generation of quarks to introduce CP violation
into the standard model and hence to explain the results of Christensen, Cronin,
Fitch and Turlay \cite{CP}. What of sterile neutrinos? How might these new 
degrees of freedom, as yet hypothetical, be fitted into particle theory?
Would their existence actually {\it explain} anything?

These are some of the questions I will explore in this lecture. I am going
to use a model-builder's perspective: starting with the standard model
and the gauge theory rule book, how might sterile neutrinos enter the game?
Quite deliberately, I will hardly address the phenomenological evidence
for sterile neutrinos, because I do not think the time is yet ripe for 
drawing definite conclusions. Following every phenomenological twist and
turn can be more a test of nimbleness than resolve! Nevertheless, I do
wish to make an observation on the currently 
favoured $2 + 2$ and $3 + 1$ scenarios \cite{2231}:
the simplest standard model extension featuring just one
sterile neutrino cannot accomodate the parameter space 
required.\footnote{This extends the material delivered in the actual lecture.}

In the next section I review
the na\"{\i}ve proposal that sterile neutrinos be
identified with ``right handed neutrinos''. A critical
discussion of the simple expedient of adding three gauge singlet fermions to
the usual minimal standard model matter content is 
followed in Sec.\ 3 by an examination of right handed neutrinos
in extended theories. Section 4 confronts the unbearable
lightness of being sterile problem: why should these
apparently alien degrees of freedom inhabit the same very low mass
range as the active neutrinos? Cosmology is briefly discussed
in Sec.\ 5, and a conclusion is then presented.

\section{Sterile neutrinos and the standard model}

Under the standard model gauge group,
\be
G_{SM} = {\rm SU}(3)_c \otimes {\rm SU}(2)_L \otimes 
{\rm U}(1)_Y,
\ee
one generation or family of quarks and leptons forms the
reducible representation,
\bea
& Q_L = \left( \begin{array}{c} u_L \\ d_L \end{array} \right)
\sim (3,2)(1/3),\qquad d_R \sim (3,1)(-2/3),\qquad u_R \sim (3,1)(4/3);&
\nn\\
& \ell_L =  \left( \begin{array}{c} \nu_L \\ e_L \end{array} \right)
\sim (1,2)(-1),\qquad e_R \sim (1,1)(-2),\qquad {\rm ?\ MISSING\ ENTRY\ ?}&
\eea
in the minimal model. A mismatch between quark and lepton degrees of
freedom is immediately evident: while the up and down quarks have
both left and right chiral components, the neutrino is purely left handed.
Bearing in mind that electric charge $Q$ is given by
\be
Q = I_L + \frac{Y}{2},
\ee
where $I_L = \sigma_3/2$ in weak isospin SU(2)$_L$ space, we see that
the ``missing'' right handed neutrino state should be
\be
\nu_R \sim (1,1)(0).
\ee
It has the quantum numbers of the vacuum, and is thus {\it sterile}
with respect to the standard model gauge interactions. The 
putative right handed neutrinos (perhaps one per family) are the most 
obvious sterile neutrino candidates. Strictly speaking, the term
``sterile fermions'' should be preferred to ``right handed neutrinos'', 
because the latter signifies the Dirac neutrino special case.
On the other hand, ``right handed neutrinos'' does better
emphasise the intrafamilial relationship between the sterile fermions 
and the other quarks and leptons.

It is interesting that the chiral structure of a quark-lepton family
provides strong motivation for one sterile fermion per family. The
presence of $\nu_R$'s would enhance two aesthetic qualities:
left-right similarity (for each left handed fermion there is
a right handed partner) and quark-lepton similarity (for each
quark of a given chirality there is an associated chiral lepton).
Historically, quark-lepton similarity was used by Bjorken and Glashow
and others to predict the existence of the charm quark \cite{charm}. Notice
that this aesthetic motivation for charm preceded the technical
motivation supplied by GIM. Aesthetics matter! One can even
upgrade these {\it similarities} into {\it symmetries}, as
per left-right symmetric models \cite{LR}, SO(10) grand 
unification \cite{so10},
and discrete quark-lepton symmetric theories \cite{ql}. There is an
exception to the rule that ``tidier families imply right handed
neutrinos'': SU(5) grand unification using the $5^* \oplus 10$
representation has none \cite{su5}.

So, let us add one right handed neutrino per family, and examine
implications for neutrino mass. As everyone knows, quark and
charged lepton mass generation is associated with spontaneous
electroweak symmetry breakdown,
\be
{\rm SU}(2)_L \otimes {\rm U}(1)_Y \to {\rm U}(1)_Q.
\ee
In the standard model, this is induced by Higgs boson 
self-interactions that lead to a nonzero vacuum 
expectation value (VEV) for 
a Higgs doublet field $\Phi$, where
\be
\Phi \sim (1,2)(1)
\ee
and
\be
\langle \Phi \rangle 
= \left( \begin{array}{c} 0 \\ v \end{array} \right).
\ee
Quark and charged lepton (Dirac) masses are produced through the
Yukawa couplings
\be
{\cal L}_{\rm Yuk} = h_d\, \overline{Q}_L\, d_R\, \Phi +
h_u\, \overline{Q}_L\, u_R\, \tilde{\Phi} +
h_e\, \overline{\ell}_L\, e_R\, \Phi + {\rm H.c.}
\ee
with $m_f = h_f v$, and where $\tilde{\Phi} \equiv i \sigma_2 \Phi^*$.
(The $h_f$'s and the $m_f$'s are $3 \times 3$ matrices in family space.)

If $\nu_R$'s are absent, and the Higgs sector remains minimal, then
the neutrinos are massless.\footnote{Just as an example:
A Higgs triplet coupling to the left handed
lepton bilinear $\overline{\ell}_L (\ell_L)^c$ would be required
to induce tree-level neutrino masses in the absence of right handed
neutrinos. The nonzero triplet VEV would also spontaneously
break lepton number, and produce a Goldstone boson called the Majoron.
To make the Majoron phenomenologically acceptable would then
require an epicyclic construction.} The individual family lepton
numbers $L_{e,\mu,\tau}$ emerge as accidental exact symmetries in
this case.

But if $\nu_R$'s exist, then neutrinos
are naturally massive. First, there are Dirac masses induced through
$\langle \Phi \rangle$:
\be
h_\nu\, \overline{\ell}_L\, \nu_R\, \tilde{\Phi} \Rightarrow
m_D = h_\nu v.
\label{eq:mDirac}
\ee
But, the trivial gauge quantum numbers of the $\nu_R$'s also
allow a bare Majorana mass matrix $M$ through
\be
{\cal L}_{\rm Maj} = M\, \overline{(\nu_R)^c}\, \nu_R + {\rm H.c.}
\label{eq:mMajorana}
\ee
Dirac mass mixing in general violates individual family
lepton number conservation but preserves total lepton number
$L = L_e + L_\mu + L_\tau$. Majorana masses and mixings violate all of
the leptonic global symmetries including $L$, with their
main phenomenological signature being neutrinoless double $\beta$-decay,
another topic covered at this School.

So, sterile neutrinos are a natural addition to the fermionic zoo of
the minimal standard model. They arguably fill a gap in the quantum
number spectrum of a family, and they generally lead to nonzero 
neutrino masses and mixings. But, the appealing see-saw mechanism \cite{seesaw}
requires them to be very massive and hence irrelevant for present
neutrino phenomenology.\footnote{Except indirectly by allowing the nonzero
neutrino masses and mixings.}

What is the see-saw mechanism and why is it considered appealing?
According to Eqs.\ \ref{eq:mDirac} and \ref{eq:mMajorana}, the
full neutrino mass matrix is
\be
\left( \begin{array}{cc}
\overline{\nu}_L & \overline{(\nu_R)^c} \end{array} \right)
\left( \begin{array}{cc} 0 & m_D \\ m_D^T & M \end{array} \right)
\left( \begin{array}{c} (\nu_L)^c \\ \nu_R \end{array} \right).
\label{massmatrix}
\ee
Now, Dirac masses in the standard model, including those
for neutrinos, are proportional to the electroweak symmetry
breaking scale $v$. So, while we do not understand the
pattern of quark and charged lepton masses revealed experimentally,
neutrino Dirac masses in a similar range are a natural expectation.
However, $M$ has a completely different origin in that it is not
proportional to $v$. Without further theoretical input, there
can be no strong prejudice about its value.

Let us specialise to just one family for ease of exposition.
The see-saw model supposes that
\be
M \gg m_D
\label{eq:seesawlimit}
\ee
so that the eigenvalues become approximately
\be
\frac{m_D^2}{M}\quad {\rm and}\quad M,
\ee
with eigenvectors
\be
\nu'_L \simeq \nu_L - \frac{m_D}{M} (\nu_R)^c,\quad
\nu'_R \simeq \nu_R + \frac{m_D}{M} (\nu_L)^c,
\ee
respectively. (Put another way, the mixing angle $\theta$
is approximately equal to $m_D/M \ll 1$.)
The predominantly sterile 
eigenstate $\nu'_R$ is {\it very massive}.

The parameter space defined by Eq.\ \ref{eq:seesawlimit} is
considered appealing because then the small eigenvalue obeys
\be
\frac{m_D^2}{M} \ll m_D \sim m_{u,d,e},
\ee
so we have a sketchy explanation for why neutrinos are much
lighter than all other known fermions. Strictly, though, this
argument just replaces the small-neutrino-mass mystery
with the large-Majorana-$\nu_R$-mass mystery. We will see 
shortly that in many extended theories, $M$ is proportional
to a high symmetry breaking scale rather than being a
bare mass. The additional theoretical assumption that there
is a symmetry breaking scale much larger than the electroweak
seems necessary to flesh out the see-saw paradigm. But one 
should acknowledge that this {\it is} an assumption, as
yet empirically unsupported. One can only hope that direct
experimental evidence for new very short-distance physics
will eventually be produced.

The limiting case opposite to that of the see-saw is also
amusing \cite{pd}. If
\be
M \ll m_D,
\label{eq:pseudoDiraclimit}
\ee
then the eigenvalues are approximately
\be
m_D \pm \frac{M}{2}
\ee
and the mixing angle is given by
\be
\tan 2\theta = - \frac{2m_D}{M} \Rightarrow 
|\theta| \simeq \frac{\pi}{4}.
\ee
This is called ``pseudo-Dirac structure'' because the
neutrino becomes fully Dirac as $M \to 0$. The signatures
for pseudo-Dirac neutrinos are:
\begin{itemize}
\item a nearly degenerate pair with a mass gap $m_D$ above zero;
\item nearly maximal active-sterile mixing.
\end{itemize}
Maximal mixing is certainly an interesting feature, both
theoretically and phenomenologically.
There is an obvious drawback, though, because in this case phenomenology
requires $m_D$ to be tiny compared to all other Dirac masses:
the fermion mass hierarchy puzzle becomes even more profound.
Nevertheless, since we do not understand the origin of this
hierarchy, tiny neutrino Dirac masses remain a possibility.
We will return to this issue in the next section.

The issue of neutrino mixing angles is just as interesting
as the origin and magnitudes of neutrino masses. So far, we have
uncovered a connection between light sterile neutrinos and large
mixing angles in the pseudo-Dirac limit. 
Later we will see that the mirror matter model supplies a
rationale for both light effectively sterile neutrinos
and large active-sterile mixing angles.
But neither the see-saw limit
nor the pseudo-Dirac scenario nor the mirror matter hypothesis
implies constraints on the
pattern of interfamily mixing. Prior to the confirmation of the
atmospheric neutrino anomaly, there was a strong theoretical
prejudice in favour of small active-active mixing angles,
simply because that was the observed situation in the quark sector.
In recent years, this prejudice has been ``revised''. Many
have made the observation that the quark and lepton sectors
need not be qualitatively similar, because of the presence
of the Majorana mass matrix $M$ for neutrinos. Since this is
a lecture on sterile neutrinos, I will refrain from
developing the active-active mixing angle story further,
except to note that a direct neutrino oscillation resolution to
the LSND anomaly \cite{lsnd} requires at least one active-active mixing
angle to be small.

The see-saw and pseudo-Dirac cases are two interesting limits. But
what can one say in general about the mass matrix of Eq.\ \ref{massmatrix}?
It is interesting that it is not an {\it arbitrary} $6 \times 6$ symmetric
matrix: the $3 \times 3$ zero matrix in the top left block ensures
that. Physical possibilities are consequently constrained.

To illustrate this, consider again just a single family. The mass
matrix
\be
\left( \begin{array}{cc} 0 & m \\ m & M \end{array} \right)
\label{onefamily}
\ee
has eigenvalues
\be
m_{\pm} = \frac{ M \pm \sqrt{ M^2 + 4 m^2 } }{2} 
\ee
and the mixing angle is given by $\tan 2\theta = -2m/M$. (The
sign of the negative eigenvalue can be absorbed into
the corresponding Majorana eigenfield.)
The zero in Eq.\ \ref{onefamily} has as an important
consequence in that the three quantities ``overall mass scale'',
``mass difference'' and ``mixing angle'' are not
arbitrary, but satisfy a relation. Defining
\bea
\Delta m^2 \equiv m_+^2 - m_-^2 & = & M\sqrt{ M^2 + 4 m^2 },\nn\\
\Sigma m^2 \equiv  m_+^2 + m_-^2 & = & M^2 + 2m^2
\eea
we can write this relation as
\be
\Sigma m^2 = \frac{1}{2} \Delta m^2 (\cos 2\theta + \sec 2\theta).
\label{eq:rhnurelation}
\ee
Within this (unrealistic one family) model, a measurement of the oscillation
parameters $\Delta m^2$ and $\theta$ would immediately specify
the absolute mass scale $\Sigma m^2$.

A more realistic scenario is to add one gauge singlet fermion to
the three standard families. This is the simplest ``three active
plus one sterile neutrino model'' imaginable. 
In the absence of Higgs triplets, the most general
mass matrix is
\be
\left( \begin{array}{cccc} 0 & 0 & 0 & m_1 \\
0 & 0 & 0 & m_2 \\
0 & 0 & 0 & m_3 \\
m_1 & m_2 & m_3 & M
\end{array} \right)
\ee
in the $[\nu_{eL}, \nu_{\mu L}, \nu_{\tau L}, (\nu_R)^c]$ basis.
The $m_i$ are Dirac masses while $M$ is the $\nu_R$ Majorana
mass. It is easy to see that this matrix has two zero eigenvalues,
so there are only two different $\Delta m^2$ parameters. This
is amusing, because it means that the currently favoured
phenomenological fits, the $2+2$ and $3+1$ so-called models,
{\it cannot} be accomodated within this minimal framework. Recall
that {\it three} unrelated $\Delta m^2$ values are required to
simultaneously resolve the solar \cite{solar,sno}, 
atmospheric \cite{atmos} and LSND \cite{lsnd}
anomalies through oscillations. So, if you want a gauge theoretic
underpinning for the aforementioned phenomenological fits,
you need to increase the number of sterile flavours or
introduce Higgs boson triplets or both \cite{bm}.

\section{``Sterile'' neutrinos beyond the standard model}

The na\"{\i}ve minimal standard model extension discussed in the previous
section sees sterile neutrinos identified with right handed
neutrinos and having the gauge quantum numbers of the vacuum.
In fact we can still sensibly call the resulting theory 
the ``standard model'', though it should no longer be called
the ``{\it minimal} standard model''. In this section,
we will consider extensions of the standard model
obtained by enlarging $G_{SM}$. It turns
out that, in most such theories, the right handed neutrinos
are {\it not} sterile with respect to the new gauge
interactions.

It may be helpful to introduce some terminology. Let us define
\begin{itemize}
\item ``fully sterile'' to mean {\it feels no gauge interactions
of any sort, including hypothetical forces beyond those of the
standard model}, and
\item ``weakly sterile'' to mean {\it does not feel standard
model gauge interactions (strong, electromagnetic or
left handed weak)}.
\end{itemize}
The following points should be noted:
\begin{itemize}
\item Right handed neutrinos in the standard model are fully sterile.
\item Full sterility is defined with respect to gauge interactions
only. Such species may well interact through Higgs boson exchange
and of course they may partake of mass mixing. They necessarily
couple via gravity.
\item As far as current neutrino phenomenology is concerned,
fully and weakly sterile neutrinos are indistinguishable. However,
there is an important theoretical difference between the two,
and phenomenological differentiation will be evident in other
contexts (such as the early universe).
\end{itemize}

Right handed neutrinos in extensions of the standard model are 
often just weakly sterile, and one can have both fully and weakly sterile states
in the same theory. Some examples are:
\begin{itemize}
\item The ``usual'' left-right symmetric model (LRSM): weakly sterile.
\item One can construct an ``unusual'' LRSM which has both weakly and fully 
sterile fermions.
\item The mirror matter or exact parity model \cite{epm}: weakly sterile or both.
\item The Pati-Salam model \cite{ps}: usually weakly sterile.
\item SU(5) grand unification: fully sterile.
\item SO(10) grand unification: usually weakly sterile.
\end{itemize}
I now expand on a couple of these examples.

\subsection{Left-right symmetric models: usual incarnation}

Left-right symmetric models are defined by the gauge group
\be
G_{\rm LR} = {\rm SU}(3)_c \otimes {\rm SU}(2)_L \otimes
{\rm SU}(2)_R \otimes {\rm U}(1)_{B-L},
\ee
with a fermionic family assigned as per
\bea
& {\rm Quarks:}\quad Q_L \sim (3,2,1)(1/3),\quad Q_R \sim (3,1,2)(1/3),&\nn\\
& {\rm Leptons:}\quad \ell_L \sim (1,2,1)(-1),\quad \ell_R \sim (1,1,2)(-1).&
\label{eq:LRSMfamily}
\eea
The basic motivation is to treat left and right handed fermions more
symmetrically than does the standard model. Parity violation is usually
induced spontaneously rather than engineered explicitly, and phenomenology
of course requires the breaking scale for right handed weak isospin SU(2)$_R$
to be suitably high (greater than a few TeV).

Different incarnations of LRSMs are defined by their Higgs sectors and
additional multiplets of fermions (if any). In the standard incarnation,
three copies of the multiplets in Eq.\ \ref{eq:LRSMfamily} specify
the complete fermion spectrum. The left handed and right handed neutrinos
reside within $\ell_L$ and $\ell_R$, respectively:
\be
\ell_{L,R} = \left( \begin{array}{c} \nu_{L,R} \\ e_{L,R} \end{array} \right).
\ee
Rather than being fully sterile, the $\nu_R$'s now participate in right handed
weak interactions, mediated by exotic $W$-like bosons and an additional neutral
gauge boson $Z'$. The right handed neutrinos are weakly sterile in the usual LRSM.

The usual LRSM is completed by specifying the Higgs sector, which is constructed
to yield a see-saw structure for neutrinos while at the same time
spontaneously breaking $G_{\rm LR}$ in two stages:
\be 
G_{\rm LR} \to G_{\rm SM} \to {\rm SU}(3)_c \otimes {\rm U}(1)_Q.
\ee
The Higgs multiplets are
\be
\Phi \sim (1,2,2)(0),\qquad \Delta_L \sim (1,3,1)(2),\qquad \Delta_R \sim
(1,1,3)(2),
\ee
which participate in the Yukawa couplings
\bea
{\cal L}_{\rm Yuk} & = & h_Q\, \overline{Q}_L\, Q_R\, \Phi + h'_Q\,
\overline{Q}_L\, Q_R\, \Phi^c
 +   h_{\ell}\, \overline{\ell}_L\, \ell_R\, \Phi + h'_\ell\,
\overline{\ell}_L\, \ell_R\, \Phi^c \nn\\
& + & \lambda_L\, \overline{(\ell_L)^c}\, \ell_L\, \Delta_L 
+ \lambda_R\, \overline{(\ell_R)^c}\, \ell_R\, \Delta_R + {\rm H.c.}
\eea
(A left-right or parity discrete symmetry is also usually imposed.)
The required VEV hierarchy
\be
\langle \Delta_R \rangle \sim v_R \gg \langle \Phi \rangle \sim v_{\rm ew}
\gg \langle \Delta_L \rangle \sim v_L
\ee
is arranged by a suitable choice of Higgs potential parameters. At the scale
$v_R$, right handed weak isospin is spontaneously broken and the
$W_R$ and the $Z'$ bosons acquire large masses. The most interesting
point for our discussion is that the same VEV generates large
Majorana masses, $\lambda_R\, v_R$, which immediately connects
the see-saw limit with an {\it a priori} separate physical phenomenon: the
spontaneous breakdown of an enlarged gauge group. There are in general
two electroweak scale VEVs within $\langle \Phi \rangle$, through which
all Dirac masses are induced via the $h$- and $h'$-terms in the Yukawa
Lagrangian. The left handed triplet VEV must be very small for
the achievement of the see-saw limit,
since the Majorana mass matrix for left handed neutrinos is proportional
to it (there is also a
phenomenological constraint from the electroweak $\rho$-parameter).

The LRSM sketched above is an example of a standard model extension that
can incorporate the see-saw mechanism. The generic lesson is that the
large right handed neutrino Majorana masses required for this
mechanism can often be correlated with a high symmetry breaking scale for
a non-standard gauge interaction that couples to
the now just weakly sterile $\nu_R$'s. This is why one often hears the claim
that neutrino oscillation phenomenology is a ``window'' into 
high-energy-scale physics. It may be or it may not be. The see-saw idea
is attractive, but it can hardly be considered as established. It
is interesting to ponder how enough experimental information could
ever be gathered to establish such a scenario beyond reasonable doubt.
The difficulty of this is just an example of the general problem of
testing theories that postulate new physics at very high energy scales. In any
case, we will obviously have to explore other paths in our search for a decent
theory of {\it light} sterile neutrinos.

\subsection{A model with both weakly and fully sterile neutrinos}

Just for amusement, let us construct a scenario featuring both weakly
and fully sterile fermions. We will adopt the gauge group of the LRSM,
but choose a different fermion and Higgs boson content. In addition
to the quarks, we have
\bea
&{\rm Leptons}:\qquad \ell_L \sim (1,2,1)(-1),\qquad 
\ell_R \sim (1,1,2)(-1);&\nn\\
&{\rm Sterile\ fermion}:\qquad S_L \sim (1,1,1)(0);&\nn\\
&{\rm Higgs\ bosons}:\qquad \Phi \sim (1,2,2)(0),\qquad 
\chi \sim (1,1,2)(1)&
\eea
We have not imposed the $L \leftrightarrow R$ discrete symmetry.
The neutrino Yukawa and bare mass terms can be assembled into
\be
\left( \begin{array}{ccc}
\overline{\nu}_L & \overline{(\nu_R)^c} & \overline{S}_L \end{array} \right)
\left( \begin{array}{ccc}
0 & \Phi & 0 \\ \Phi & 0 & \chi \\ 0 & \chi & M_S \end{array} \right)
\left( \begin{array}{c} (\nu_L)^c \\ \nu_R \\ (S_L)^c \end{array} \right),
\ee
where I am being schematic rather than technically accurate.
The $\nu_R$ is weakly sterile, whereas $S_L$ is fully sterile. 

A VEV for Higgs multiplet $\chi$ is required to break right handed weak isospin
at a high scale, while $\langle \Phi \rangle$ sets the electroweak scale.
The fully sterile fermion has a bare Majorana mass $M_S$.
For $M_S \ll \langle \Phi \rangle \ll \langle \chi \rangle$, 
the lightest eigenstate is a Majorana
fermion of mass $M_S \langle \Phi \rangle^2/\langle \chi \rangle^2$ 
which is predominantly $\nu_L$.
The other two eigenstates (weakly and 
fully sterile) are of order $\langle \chi \rangle$ and thus very massive.

This scenario illustrates that there can 
in principle be a ``hierarchy of sterility'' for neutrino-like particles,
but in the above model neither of the sterile states is light.

\subsection{The story so far}

Let us pause to summarise what we have deduced so far:
\begin{itemize}
\item There are varieties of ``sterile'' neutino, grouped into the broad
categories of weakly and fully sterile.
\item Their existence can be very well motivated by quark-lepton
and left-right similarity (or symmetry, if you want to go that far).
\item The pseudo-Dirac limit hints at a connection between large
mixing angles and sterile neutrinos.
\item {\it The main issue is whether they are expected to be heavy
or light. The see-saw mechanism favours heavy sterile neutrinos.}
Such particles are usually called ``heavy neutral leptons'',
and while they may have an important role to play in cosmological
baryogenesis (a topic beyond the scope of this lecture), they
play no direct role in neutrino oscillation phenomenology.
\item The model obtained by adding a single sterile fermion to
the minimal standard model cannot serve as a gauge theoretic
underpinning for the presently favoured $2+2$ and $3+1$
phenomenological fits.
\end{itemize}

\section{The unbearable lightness of being sterile}

Let us now confront what has emerged as a core issue: can one 
theoretically justify {\it light} sterile neutrinos?

\subsection{Naturalness and technical naturalness}

Clearly, one can just add gauge singlet fermions 
with arbitrary masses to any model, so what is the problem?
It is considered a question of {\it naturalness} rather
than mere possibility. (And one has to worry about how
gauge singlet fermions couple to the other degrees of
freedom.)

``Naturalness'' is an aesthetic concept, and physicists
can have different opinions about what it means.
``Technical naturalness'' is a precise mathematical criterion,
which is usually weaker than naturalness per se. A parameter
choice is {\it technically natural} if its adoption increases
the symmetry of the theory. (Parameter choice means either a
special numerical value, or a special relationship between
parameters.) Increased symmetry ensures that the
special parameter choice is not altered by radiative corrections.
Actually, a slightly weaker statement is more pertinent: 
Consider a parameter $\lambda$, and suppose that for 
$\lambda = \lambda_0$, the symmetry of the theory is increased.
Radiative corrections cannot move $\lambda$ from having the
value $\lambda_0$. But then one also deduces that
values of $\lambda$ in the neighbourhood of $\lambda_0$
enjoy a kind of stability, because radiative
corrections to the parameter must be proportional
to the difference $\lambda - \lambda_0$, which 
by hypothesis has a small value. Since the radiative
corrections are thus also small, points in the
neighbourhood of $\lambda_0$ never move out of that regime.
Such a parameter choice is also termed technically natural.

Well, taking the mass of a sterile neutrino
to zero is a special parameter choice. If technical
naturalness holds, then massless sterile neutrinos
remain massless to all orders. 

For definiteness, consider
the standard model with right handed neutrinos added. Taking
the $\nu_R$ Majorana mass to zero increases the symmetry of the
theory, because total lepton number conservation then results.
So, having Dirac rather than Majorana neutrinos is
technically natural. But is it natural? What do {\it you}
think? Many would say ``No!'', because it is nicer to
include all renormalisable and gauge invariant terms in
the Lagrangian. In response, one might add small Majorana
masses. Then, technically, they will
remain small to all orders in perturbation theory.
This remark is obviously relevant for the pseudo-Dirac
option.

Taking in addition the neutrino Dirac mass $m_D$ to
zero increases the symmetry further, because the chiral
transformation,
\be
\nu_R \to e^{i\alpha}\, \nu_R,
\ee
with every other field just going into itself, 
is now an invariance. Indeed, this is just an instance
of the well known fact that zero fermion masses go
hand-in-hand with increased chiral symmetry.

Thus we conclude that having light sterile (and active!)
neutrinos is technically natural.\footnote{Actually, one can
question if the concept of technical naturalness is truly
meaningful. The point is that it presupposes a perturbative
analysis: One chooses a special parameter choice at
tree-level. One calculates 1-loop corrections
and asks if the special choice still holds. It will if
the choice is associated with an increase in symmetry, be it
exact or approximate. Otherwise it probably will not 
hold in general, and
one may then have to fine-tune to maintain the special value
after 1-loop renormalisation. One then calculates 2-loop
corrections, and repeats the process. And so on. 
Technically natural choices display perturbative stability,
but why should stability with respect to a certain
approximation scheme be so fundamentally important?
This is an interesting question in light of the usual
motivation for supersymmetry as well. Even if one accepts
this criticism, parameter
space regions near points of enhanced symmetry
are still mathematically special compared to generic regions.}
But is it natural, is it nice?
Again, many would say ``No!'', typical opinions being:
\begin{itemize}
\item Small Majorana masses are not nice, because in many
extensions of the standard model they are proportional to
a higher symmetry breaking scale (as we have seen).
(Such an opinion contains the implicit assumption that
additional symmetry breaking scales are desirable and/or
are likely to exist.)
\item Small Dirac masses are not nice, because they
are proportional to the electroweak VEV, so you would
need an extremely small Yukawa coupling constant.
\end{itemize}
In response to such criticisms, one can try to invent
models that purport to explain why Majorana and Dirac
masses should be so small. This is either a model-building
challenge, or an epicyclic indulgence, depending on your
point of view.

\subsection{Mirror neutrinos}

There is no known model for fully sterile light neutrinos that is
generally accepted as being ``nice''. I should comment that
the right handed neutrino identification is not the only
possible one, especially in supersymmetric theories
which abound with states such as axinos and modulinos
that are sterile ``neutrino''
candidates. However, I will concentrate on
a completely different possibility: mirror neutrinos.

The mirror matter or exact parity model is essentially the standard
model squared \cite{epm}. The gauge group is
\be
G  = G_{\rm SM} \otimes G'_{\rm SM},
\ee
where both factors are isomorphic to
SU(3)$\otimes$SU(2)$\otimes$U(1). Ordinary particles transform
non-trivially under the first factor and are singlets under the second.
This means that ordinary particles interact amongst themselves
in the standard way. The extension comes from postulating
a new sector called ``mirror matter''. In addition to the
gauge bosons of $G'_{\rm SM}$, new fermions and Higgs bosons are added which
transform non-trivially under the second factor but trivially under
the first. If a given ordinary fermion $f_L$ transforms as
$(R,1)$, then its mirror partner $f'_R$ transforms as $(1,R)$.
Furthermore a discrete non-standard 
parity symmetry, 
\be
f_L \leftrightarrow f'_R,
\ee
is imposed on the Lagrangian. Mirror particles interact amongst
{\it them}selves through $G'_{\rm SM}$ gauge forces that
have the same form and strength as their ordinary counterparts.
The only difference is that mirror weak interactions
are right handed, to offset the left handed nature of the ordinary weak
interactions. Our original motivation for this scheme was
to demonstrate that nature could be invariant under improper
Lorentz transformations despite the left handed nature of
the weak force. (We found out later that Lee and Yang
had sketched the mirror matter idea in their famous paper
proposing parity violation \cite{ly}!)

The mirror matter model is interesting for neutrino physics
because the mirror neutrinos are, first of all, sterile with
respect to ordinary weak interactions and, secondly,
guaranteed to be light. Mirror neutrinos are weakly sterile.
Let us have a look at this in more detail. Under
$G_{\rm SM} \otimes G'_{\rm SM}$ the ordinary lepton
doublets transform as per
\be
\ell_L = \left( \begin{array}{c} \nu_L \\ e_L \end{array}\right)
\sim [\, (1,2)(-1)\, ;\, (1,1)(0)\, ],
\ee
using a slightly cumbersome but obvious notation, whereas
the mirror lepton doublet behaves according to
\be
\ell'_R = \left( \begin{array}{c} \nu'_R \\ e'_R \end{array}\right)
\sim [\, (1,1)(0)\, ;\, (1,2)(-1)\, ].
\ee
The $\nu'_R$ fields are the mirror neutrinos. Notice that they
are completely different states from what we have been calling ``right
handed neutrinos''. Indeed, the {\it minimal} mirror matter model
does not contain right handed neutrinos (nor their parity
partners, the left handed mirror neutrinos); both ordinary
and mirror neutrinos are then exactly massless. Why are
mirror neutrinos massless? For exactly the same reasons that
ordinary neutrinos are massless: absence of the ``missing''
singlet fermion per (ordinary or mirror) family, and absence
of Higgs triplets. The discrete parity symmetry ensures
that the physics of the ordinary sector is replicated by
the mirror sector.

The fact that mirror neutrinos are weakly sterile and massless
in the minimal mirror matter model provides a good starting
point for developing our coveted 
theory for {\it light}, effectively
sterile neutrinos. All we need to do is extend the standard
sector so that ordinary neutrinos get tiny masses. Whatever
mechanism we use to achieve this (e.g.\ see-saw) will operate
analogously in the mirror sector! If ordinary neutrinos are
light, then so also will be the mirror neutrinos. In fact, we 
do not have to understand exactly why ordinary neutrinos
are light in order to conclude that mirror neutrinos
must also be light! Whatever the reason, it will have its
mirror analogue.

Let us briefly discuss the see-saw route by way of 
example. In addition to the ordinary and mirror lepton
doublets, we add a singlet fermion to each family. As
mentioned above, the singlet added to an ordinary
family is just a right handed neutrino $\nu_R$, whereas the
mirror version is called a left handed mirror neutrino
$\nu'_L$. We now write down all renormalisable Yukawa
coupling and bare mass terms consistent with the
gauge symmetry $G_{\rm SM} \otimes G'_{\rm SM}$,
and we take the see-saw limit. Switch off inter-family
mixing for simplcity. It is easy to see that
the light eigenstate sector then consists of two states
per family, which
are maximal mixtures of ordinary and mirror
neutrinos:
\be
\nu_{\pm} = \frac{\nu \pm \nu'}{\sqrt{2}}.
\label{eq:maxmix}
\ee
The state $\nu$ is mostly the $\nu_L$, while the mirror
state $\nu'$ is mostly the antiparticle of $\nu'_R$.
(This is the left handed mirror antineutrino. In the
literature, it is usually called a mirror neutrino for simplicity, even
though strictly speaking it is an antiparticle.)
The masses $m_{\pm}$ are arbitrary, except for the qualitative
constraint that they are both small due to the see-saw.
The appearence of pairwise ordinary-mirror maximal mixing
is reminiscent of the pseudo-Dirac option discussed
earlier. In this case, however, maximal mixing is enforced by
the exact discrete parity symmetry (and the lightness of the
effectively sterile state has an elegant theoretical explanation). 
Recall from basic quantum 
mechanics that symmetry eigenstates must also be Hamiltonian
eigenstates. The maximal linear combinations 
$\nu_{\pm}$ are exactly the even and
odd parity eigenstates, respectively. The mixing between
ordinary and mirror neutrinos arises from some of the
Lagrangian terms containing the singlet species, for example
\be
\overline{\ell}_L (\nu'_L)^c \tilde{\phi}
+ \overline{\ell}'_R (\nu_R)^c \tilde{\phi}',
\ee
plus other mixed terms involving the singlets only,
where $\phi'$ is the mirror Higgs doublet. The result
in Eq.\ \ref{eq:maxmix} is model independent though:
every neutrino mass model must produce that result,
provided that any additional neutrino-like states
are made sufficiently massive. Notice that the
strength of the ordinary-mirror neutrino mixing
is governed by the mass splitting $m_+ - m_-$,
since the mixing angle is constrained to be maximal.
In terms of neutrino phenomenology, this means
that the wavelength is a free parameter but the amplitude
is not.

Maximal mixing is interesting because of both the atmospheric
and solar neutrino problems. Many non-trivial experimental
results are consistent with maximal $\nu_{\mu} 
\leftrightarrow \nu'_{\mu}$ and maximal $\nu_e 
\leftrightarrow \nu'_e$ oscillations. But, the
SuperKamiokande collaboration claim that $\nu_{\mu} \to \nu_{\tau}$
is preferred 
over the sterile channel by the atmospheric neutrino data, and
the combined SuperKamiokande and SNO data \cite{solar,sno}
suggest a $3\sigma$ preference for $\nu_e \to \nu_{\mu,\tau}$
over $\nu_e \to \nu_s$. This is disappointing from
the mirror neutrino perspective, and it will be interesting
to see if future data will support these initial findings.

In closing the mirror neutrino discussion, let us
compare the effective $2 \times 2$ mass matrix for the light
ordinary plus mirror one-family situation with the
alternative $\nu_L$ plus $\nu_R$ scenario of Eq. \ref{onefamily}.
For the mirror case, the analogous matrix must be of the form
\be
\left( \begin{array}{cc} m_1 & m_2 \\ m_2 & m_1 \end{array} 
\right)
\ee
due to the discrete symmetry. As before, there are just two
parameters to describe three quantities: 
mass splitting, overall mass scale and mixing angle. In this case,
the mixing angle is uniquely singled out as the
constrained parameter, since it must be $\pi/4$, whereas
the previous case resulted in an algebraic relation, 
Eq.\ \ref{eq:rhnurelation},
involving all three of the quantities. Mirror neutrinos
are in general distinguishable from right handed neutrinos.

\section{Cosmology}

Light sterile neutrinos can have important cosmological implications.
The most dramatic possible effect concerns big bang nucleosynthesis
(BBN), the processes thought to be responsible for generating
the light isotopes $^4$He, $^3$He, D and $^7$Li. This is a rather
complicated topic, that I can only summarise here.

The BBN epoch occurs shortly after neutrinos thermally decouple from
the $e^{\pm}/\gamma$ plasma at about $T \simeq 1$ MeV, where $T$ is
temperature. The plasma contains some nucleonic contamination,
with neutrons and protons being interconverted through the
processes
\be
\nu_e n \leftrightarrow e^- p,\qquad \overline{\nu}_e p \leftrightarrow
e^+ n,\quad n \leftrightarrow p e^- \overline{\nu}_e.
\label{eq:npreactions}
\ee
These contaminants form the raw material for nucleosynthesis. The
most abundantly produced isotopes are H (just unsynthesised protons)
and $^4$He, the latter being a tightly bound nucleus. The relative
abundance of neutrons to protons essentially determines the relative
yield of $^4$He to H, since almost all of the neutrons eventually
get incorporated into $^4$He. This ratio is determined by the
relative rates of the reactions in Eq.\ \ref{eq:npreactions}
as well as by the expansion rate of the universe during the relevant
period. Light sterile neutrinos can alter the course of BBN via both of
these avenues.

The expansion of the universe during BBN
is determined by the relativistic component of the plasma. In standard
BBN, the relativistic species are the three active neutrinos and antineutrinos,
electrons and positrons, and photons. A significant light sterile neutrino
component would increase the expansion rate relative to the standard
value. This would bring forward ``weak freeze out'', the time when the reactions
of Eq.\ \ref{eq:npreactions} cease maintaining the $n/p$ ratio
at its equilibrium value, which for the
zero chemical potential case is $\exp[(m_p - m_n)/T]$. If weak freeze-out
occurs earlier, then $T$ is larger, and hence $n/p$ is also larger. This
can increase the $^4$He yield unacceptably.

An interesting set of unknown cosmological parameters are the
neutrino-antineutrino number density asymmetries for each flavour.
In standard BBN, these are put for simplicity to zero. However,
the relic neutrino background has never been detected, 
so we have no direct empirical justification for this simplifying
assumption. It turns out that active-sterile neutrino and
antineutrino oscillations can induce large asymmetries in the
active flavours \cite{ftv}.\footnote{And in the sterile flavour(s).}
Two important consequences flow from this. First, an
asymmetry in the $e$-like neutrino flavour will directly
affect the rates for Eq.\ \ref{eq:npreactions}, and thus
also $n/p$ \cite{mue}.\footnote{In equilibrium, neutrino asymmetries
are synonymous with nonzero neutrino chemical potentials.}
Second, neutrino asymmetries generate effective 
Wolfenstein potentials \cite{wolf} for active-sterile neutrino oscillation
modes, and hence act to suppress them. This is important, because
oscillations into sterile species can populate the plasma
with an extra relativistic component.

So, what can sterile neutrino cosmology look like? The physics
is quite interesting and some aspects of it are even subtle.
Relevant issues are:
\begin{itemize}
\item {\it Sterile neutrino decoupling temperature.} At some high
temperature, sterile neutrinos decouple from the rest of the plasma.
The temperature is high, because by definition sterile neutrinos
interact very weakly with all other particles. Exactly how high
depends on the precise model, on what other interactions the
``sterile'' neutrinos feel, on whether they are fully or weakly
sterile. (Note that the mirror matter model is a case unto itself,
because of the self-interactions within the mirror sector \cite{mirrorcosmo}.)
\item {\it Dilution of sterile component.} As the universe cools,
species mutually annihilate and reheat the main component of
the plasma. Since the sterile neutrinos have decoupled, they
do not get reheated and their number density becomes negligible
compared to the reheated species.
\item {\it Repopulation through oscillations?} However, 
the sterile species can make a ``comeback'' through
active-sterile neutrino oscillations. For an interesting
range of oscillation parameters, this becomes an issue in the
epoch immediately prior to BBN. For some parameter ranges, the
sterile neutrinos will be repopulated in the plasma,
entailing a higher expansion rate during BBN and thus potentially
leading to $^4$He overproduction.
\item {\it Large neutrino asymmetries?} However, if large enough
neutrino asymmetries exist in the plasma, then the active-sterile
oscillation modes will be suppressed \cite{prl}. Acceptable cosmology despite
the existence of light sterile neutrinos can result.
\item {\it Oscillation generated asymmetries.} Remarkably,
large neutrino asymmetries will be generated by the active-sterile
oscillations themselves, provided the oscillation parameters are
in the correct regime \cite{ftv}! Depending on the model and the parameters,
different asymmetry values can be produced for the different 
active flavours, including the $e$-like flavour which affects
the $n \leftrightarrow p$ reactions directly.
\end{itemize}
Cosmology with light sterile neutrinos requires careful analysis,
with the outcome depending on the model and on the actual values
of the parameters. Successful cosmologies can result.

\section{Conclusion}

``Sterile neutrino'' is a class of fermions whose
place in nature has yet to be finalised. While
acknowledging that an oscillation resolution for the combined
atmospheric, solar and LSND problems requires at least
one light sterile neutrino, I have taken my cues from
theory rather than phenomenology, describing some 
model building perspectives on how such states may fit
into the standard model or extensions thereof.
A particular concern was developing theoretical 
frameworks for how such particles could have tiny masses.

We have drawn the following conclusions:
\begin{itemize}
\item The ``missing entry'' in the minimal standard model
family is the right handed neutrino. It is an obvious
sterile neutrino candidate. Its existence would in a sense
balance out each family, by enhancing both left-right and
quark-lepton similarity. Three sterile neutrino flavours
would be implied.
\item However, there is no really compelling reason to
give such particles very small masses. In fact, the
see-saw mechanism for producing light active neutrinos
proposes that right handed neutrinos are very massive
Majorana fermions. Such species may play an important
role in cosmology, but would not be directly relevant
for neutrino oscillation phenomenology. The Devil's Advocate
points out, though, that
assigning $\nu_R$'s small masses is a technically natural
procedure.
\item The right handed neutrino paradigm in the absence
of Higgs triplets produces a restricted Majorana mass
matrix. Not all phenomenological parameter regimes
are possible, and in particular, the currently
favoured $2+2$ and $3+1$ phenomenological fits
cannot be accomodated by the minimal standard model
augmented by a single sterile fermion.
\item The opposite limit to the see-saw produces pseudo-Dirac
neutrinos, featuring maximal active-sterile mixing. However,
there is no good understanding for why such states should
be light, despite having noted that technical naturalness is satisfied.
\item One has to be careful about the meaning of ``sterile''.
The strict definition requires sterility with respect to
the standard interactions, but leaves open the possibility
of nonzero influence under hypothetical very weak forces such as
right handed weak interactions. The terminology
``fully sterile'' and ``weakly sterile'' was introduced
to deal with this.
\item Mirror neutrinos are a radical (?) reinterpretation
of the theoretical origin of sterile neutrinos. The
lightness of being problem is solved, and pairwise 
ordinary-mirror mixing is predicted. 
\item Cosmology with sterile neutrinos is subtle, interesting
and complicated, with acceptable cosmological outcomes quite
possible.
\end{itemize}

So, let us return to the questions posed in the first paragraph.
What of sterile neutrinos? How might these new 
degrees of freedom, as yet hypothetical, be fitted into particle theory?
Would their existence actually {\it explain} anything? We have seen that
they might signal a completion of a quark-lepton family. On
the other hand, they might be the first indication for a mirror sector.
Pairwise active-sterile maximal mixing can be explained by either
the pseudo-Dirac configuration or the mirror matter hypothesis.
Current phenomenological indications might lead one to be 
pessimistic about
this elegant explanation for large mixing, but perhaps all
hope is not yet lost.

In any case, the most urgent task is to perform new experiments
so as to either establish the existence of light sterile neutrinos,
or to make them clearly irrelevant. The discovery of light sterile
neutrinos would be epochal, simply because they would be genuinely
new degrees of freedom. If the mirror matter idea is correct, then
they would also be the tip of the iceberg. In addition, light sterile
neutrinos would make early universe cosmology very interesting indeed.
Here's hoping that in the future these particles enrich real
physics and not just our imaginations.

\vspace{1cm}

\noindent
Many thanks to Professor Amand Faessler for inviting me to this
School and to Jan Kuckei for his organisational assistance. 
I would also like to thank my long-time collaborator Robert Foot
for great physics, interesting times and provocative discussions. 
I remain eternally grateful to my 
neutrino-oriented (former or present) 
PhD students Nicole Bell, Roland Crocker and
Yvonne Wong for providing many insights. Finally, 
thanks are also due to my Roman
colleagues Pasquale di Bari, Paolo Lipari and Maurizio Lusignoli
for being interested in sterile neutrino cosmology (and for
giving me excuses to visit Rome). This work was supported by the
Australian Research Council.


\begin{thebibliography}{99}
\itemsep -2pt

\bibitem{GIM}
S. L. Glashow, J. Iliopoulos and L. Maiani, \Journal{\PREV}
{D2}{1285}{1970}

\bibitem{KM}
M. Kobayashi and T. Maskawa, \Journal{\PRO}{49}{652}{1973}

\bibitem{CP}
J. H. Christenson, J. W. Cronin, V. L. Fitch and R. Turlay,
\Journal{\PRL}{13}{138}{1964}

\bibitem{2231}
For recent phenomenological analyses see, for instance,
M. C. Gonzalez-Garcia, M. Maltoni and C. Pena-Garay, hep-ph/0108073;
M. Maltoni, T. Schwetz and J. W. F. Valle, \Journal{\PLB}{518}{252}{2001}

\bibitem{charm}
J. D. Bjorken and S. L. Glashow, \Journal{\PL}{11}{255}{1964}

\bibitem{LR}
J. C. Pati and A. Salam, \Journal{\PRD}{10}{275}{1974};
R. N. Mohapatra and J. C. Pati, \Journal{\PRD}{11}{566}{1975};
G. Senjanovic and R. N. Mohapatra, \Journal{\PRD}{12}{1502}{1975}

\bibitem{so10}
H. Georgi, in {\it Proceedings of the American Institute of Physics},
ed.\ C. E. Carlson (1974); H. Fritzsch and P. Minkowski, 
\Journal{\ANNP}{93}{193}{1975}

\bibitem{ql}
R. Foot and H. Lew, \Journal{\PRD}{41}{3502}{1990}; 
\Journal{\NCA}{104}{167}{1991};
R. Foot, H. Lew and R. R. Volkas, \Journal{\PRD}{44}{1531}{1991};
R. Foot and R. R. Volkas, \Journal{\PLB}{358}{318}{1995}

\bibitem{su5}
H. Georgi and S. L. Glashow, \Journal{\PRL}{32}{438}{1974}

\bibitem{seesaw}
M. Gell-Mann, P. Ramond and R. Slansky, in {\it Supergravity},
Proceedings of the Workshop, Stony Brook, New York, 1979,
eds.\ P. van Nieuwenhuizen and D. Z. Freedman (North-Holland,
Amsterdam, 1979);
T. Yanagida, in {\it Proceedings of the Workshop on Unified
Theory and Baryon Number of the Universe}, Tsukuba, Japan,
1979, eds.\ A. Sawada and A. Sugamoto; see also,
R. N. Mohapatra and G. Senjanovic, \Journal{\PRL}{44}{912}{1980}

\bibitem{pd}
See, for example,
S. M. Bilenky and B. M. Pontecorvo, {\it Sov. J. Nucl. Phys.} 38 (1983) 248;
M. Kobayashi, C. S. Lim and M. M. Nojiri, \Journal{\PRL}{67}{1685}{1991};
H. Minakata and H. Nunokawa, \Journal{\PRD}{45}{3316}{1992};
S. M. Bilenky and S. T. Petcov, \Journal{\RMP}{59}{671}{1987};
C. Giunti, C. M. Kim and V. W. Lee, \Journal{\PRD}{46}{3034}{1992};
J. P. Bowes and R. R. Volkas, {\it J. Phys. G} 24 (1998) 1249;
A. Geiser, \Journal{\PLB}{444}{358}{1999};
D. Chang and O. C. W. Kong, \Journal{\PLB}{477}{416}{2000}


\bibitem{lsnd}
C. Athanassopoulos et al., LSND Collaboration, \Journal{\PRL}{75}{2650}{1995};
77 (1996) 3082; 81 (1998) 1774


\bibitem{solar}
Y. Fukuda et al., SuperKamiokande Collaboration, \Journal{\PRL}{81}{1158}{1998};
81 (1998) 4279(E); 82 (1999) 1810; hep-ex/0103032;
J. N. Abdurashitov et al., SAGE Collaboration, {\it Phys. Atom. Nucl.} 63 (2000) 943;
W. Hampel et al., GALLEX Collaboration, \Journal{\PLB}{447}{127}{1999};
B. T. Cleveland et al., Homestake Collaboration, {\it Ap. J.} 496 (1998) 505;
M. Altmann et al., GNO Collaboration, \Journal{\PLB}{490}{16}{2000};
Y. Fukuda et al., Kamiokande Collaboration, \Journal{\PRL}{77}{1683}{1996}

\bibitem{sno}
Q. R. Ahmed et al., SNO Collaboration, \Journal{\PRL}{87}{071301}{2001}

\bibitem{atmos}
Y. Fukuda et al., SuperKamiokande Collaboration, \Journal{\PRL}{82}{2644}{1999};
Y. Fukuda et al., Kamiokande Collaboration, \Journal{\PLB}{335}{237}{1994};
W. A. Mann (for the Soudan 2 Collaboration), {\it Nucl. Phys. Proc. Suppl.}
91 (2000) 134; R. Becker-Szendy et al.,  {\it Nucl. Phys. Proc. Suppl.}
38 (1995) 331; M. Ambrosio et al., MACRO Collaboration, \Journal{\PLB}{478}{5}{2000}

\bibitem{bm}
See, for example, K. S. Babu and R. N. Mohapatra, hep-ph/0110243

\bibitem{epm}
R. Foot, H. Lew and R. R. Volkas, \Journal{\PLB}{272}{67}{1991};
{\it Mod. Phys. Lett.} A 7 (1992) 2567; R. Foot,
{\it Mod. Phys. Lett.} A 9 (1994) 169; R. Foot and R. R. Volkas,
\Journal{\PRD}{52}{6595.}{1995} Mirror matter models
with spontaneously broken parity have also been discussed:
R. Foot and H. Lew, hep-ph/9411390; Z. G. Berezhiani and
R. N. Mohapatra, \Journal{\PRD}{52}{6607}{1995};
R. Foot, H. Lew and R. R. Volkas, {\it JHEP} 0007 (2000) 032

\bibitem{ps}
J. C. Pati and A. Salam, \Journal{\PRD}{8}{1240}{1973}

\bibitem{ly}
T. D. Lee and C. N. Yang, \Journal{\PREV}{104}{254}{1956}

\bibitem{ftv}
R. Foot, M. J. Thomson and R. R. Volkas, \Journal{\PRD}{53}{5349}{1996};
R. Foot and R. R. Volkas, \Journal{\PRD}{55}{5147}{1997}.
For background and other reading see, for example: L. Stodolsky, \Journal{\PRD}{36}{2273}{1987};
A. Dolgov, {\it Sov. J. Nucl. Phys.} 33 (1981) 700; K. Enqvist, K. Kainulainen
and J. Maalampi, \Journal{\NPB}{349}{754}{1991}; B. H. J. McKellar
and M. J. Thomson, \Journal{\PRD}{49}{2710}{1994};
P. di Bari, P. Lipari and M. Lusignoli, {\it Int. J. Mod. Phys.} A 15 (2000) 2289;
R. R. Volkas and Y. Y. Y. Wong, \Journal{\PRD}{62}{093025}{2000}.
For a review
and a more complete list of references on active-sterile neutrino
oscillations and cosmology see: M. Prakash, J. M. Lattimer, R. F. Sawyer
and R. R. Volkas, astro-ph/0103095, {\it Ann. Rev. Nucl. Part. Sci.} 51 (in press)


\bibitem{mue}
R. Foot and R. R. Volkas, \Journal{\PRD}{56}{6653}{1997}; D 59 (1999) 02991(E);
N. F. Bell, R. Foot and R. R. Volkas, \Journal{\PRD}{58}{105010}{1998}.
There have been some controversies in this field; see, for example:
X. Shi, G. M. Fuller and K. Abazajian, \Journal{\PRD}{60}{063002}{1999} and
R. Foot, \Journal{\PRD}{61}{023516}{1999}


\bibitem{wolf}
L. Wolfenstein, \Journal{\PRD}{17}{2369}{1978}; D 20 (1979) 2634;
S. P. Mikheyev and A. Yu. Smirnov, {\it Nuovo Cimento} C 9 (1986) 17


\bibitem{mirrorcosmo}
R. Foot and R. R. Volkas, \Journal{\PRD}{61}{043507}{2000}

\bibitem{prl}
R. Foot and R. R. Volkas, \Journal{\PRL}{75}{4350}{1995}


\end{thebibliography}
\end{document}